# All-Optical Domain Inversion in LiNbO$_3$ Crystals by Visible Continuous-Wave Laser Irradiation

*Carlos Sebastián-Vicente, Jörg Imbrock, Simon Laubrock, Olga Caballero-Calero, Angel García-Cabañes, and Mercedes Carrascosa\**

C. Sebastián-Vicente, A. García-Cabañes, M. Carrascosa
Departamento de Física de Materiales, Universidad Autónoma de Madrid, 28049 Madrid, Spain
E-mail: m.carrascosa@uam.es

C. Sebastián-Vicente, A. García-Cabañes, M. Carrascosa
Instituto Nicolás Cabrera, Universidad Autónoma de Madrid, 28049 Madrid, Spain

J. Imbrock, S. Laubrock
Institute of Applied Physics, University of Münster, Corrensstr. 2, 48149 Münster, Germany

O. Caballero-Calero
Instituto de Micro y Nanotecnología, IMN-CNM, CSIC (CEI UAM+CSIC) Isaac Newton, 8, Tres Cantos, E-28760 Madrid, Spain

**Abstract**

LiNbO$_3$ is a distinguished multifunctional material where ferroelectric domain engineering is of paramount importance. This degree of freedom of the spontaneous polarization remarkably enhances the applicability of LiNbO$_3$, for instance in nonlinear photonics. However, conventional electrical poling suffers from several drawbacks. Namely, the lithographic patterning of electrodes is cumbersome and expensive, and the spatial resolution is constrained. In this work, we report the first method for all-optical domain inversion of LiNbO$_3$ crystals using continuous-wave visible light. While we mainly focus on iron-doped LiNbO$_3$, the applicability of the method is also showcased in undoped congruent LiNbO$_3$. The technique is outstandingly simple, low-cost and readily accessible. It relies on ubiquitous elements: a light source with low/moderate power, basic optics and a conductive surrounding medium, e.g. water. Light-induced domain inversion is unequivocally demonstrated and characterized by combination of several experimental techniques: pyroelectric trapping of charged microparticles, selective chemical etching, surface topography profilometry, scanning electron microscopy and 3D Čerenkov microscopy. The influence of light intensity, exposure time, laser spot size and surrounding medium is thoroughly studied. Overall, our all-optical method offers straightforward implementation of LiNbO$_3$ ferroelectric domain engineering, potentially sparking new research endeavors aimed at novel optoelectronic applications of photovoltaic LiNbO$_3$ platforms.

## 1. Introduction

Over many decades, lithium niobate (LN) has been a major playground for photonics research due to the rich variety of properties combined in a single material platform. LN stands out for its high electrooptic coefficients with fast response, high nonlinear coefficients for second-order $\chi^{(2)}$ and third-order $\chi^{(3)}$ processes, strong photorefraction and broad transparency window (350-5500 nm).[1,2] These remarkable properties make it possible to tailor this platform for a number of widespread photonic applications: electrooptic modulators,[3,4] integrated spectrometers,[5] hologram storage,[6] frequency conversion,[7,8] frequency comb generation,[9,10] optofluidic lab-on-a-chip devices[11,12] or various nonlinear nanophotonic applications,[13] among others.

Most of the aforementioned applications of LN are ultimately enabled by its non-centrosymmetric and ferroelectric nature. Indeed, LN has a remarkably high Curie temperature of 1210 °C, showing a ferroelectric polar phase in a broad temperature range. Due to the lack of inversion symmetry of the lattice in the polar phase, LN also exhibits the piezoelectric, pyroelectric and bulk photovoltaic (PV) effects, which are particularly strong in this material.[14] Altogether, these effects allow the efficient transduction of mechanical, thermal or optical stimuli into tailored electric fields. This provides even more degrees of freedom to LN, both inside and outside the realm of photonics. For instance, acousto-optic modulators,[15] photodetectors,[16] manipulation of microfluidic droplets[17–19] or patterning of micro/nanoparticles,[20,21] are some successful examples exploiting these effects. In the framework of ferroelectricity, some of the applications of LN require, or can benefit from, customized ferroelectric domain structures.[22–24] The most classical example is frequency conversion by second-harmonic generation (SHG), where periodically-poled structures are needed to achieve quasi-phase matching.[7] However, domain-engineered structures in LN are also of great interest for surface acoustic wave devices,[25] conductive domain walls for nanoelectronics,[26] trapping and patterning of particles,[13,20,21,24] and many more.

The traditional method for domain inversion consists in the application of external electric fields above the coercive field ($\sim$210 kV cm$^{-1}$ in congruent undoped LN)[22] in the opposite direction of the spontaneous polarization. To achieve the desired domain patterns, electrodes are patterned on the crystal by lithography techniques, thus making the process cumbersome and costly. Moreover, the fabrication of domain structures with submicron resolution is hindered by the lateral growth of the inverted domains beyond the extent of the patterned electrodes. As an alternative, other domain inversion methods have been proposed, such as electron beam irradiation,[27] ion beam irradiation[28] or light-mediated poling.[29] While electron/ion beams are capable of high resolutions, they require expensive and complex equipment, they are constrained to shallow domains and lack uniformity for large-area patterning for industrial usage.[22] On the other hand, certain light-based techniques are promising candidates that can overcome some of the limitations of these methods, taking advantage of the intrinsic patterning capabilities of light.

Optical domain engineering falls into two major groups: light-assisted poling and all-optical poling.[29] While all-optical techniques purely rely on light itself, light-assisted methods still require an external electric field[30–33] or additional processing steps, such as film patterning by electron beam lithography[34] or pyroelectric thermal treatments.[35,36] All-optical methods completely circumvent the need for additional electric fields or processing steps. All-optical poling can be sub-classified as a function of the excitation wavelength:

- UV light (pulsed or continuous-wave CW).[37–39]
- Far-IR light at $\lambda$ = 10.6 µm (pulsed).[40]
- Near-IR light (pulsed).[41–45]

The first two methods share that they work outside the transparency window of LN, either with above-bandgap UV light or IR light beyond 5.5 µm. Thus, the absorption is remarkably high, leading to thermally-induced electric fields due to local temperature variations. Different contributing mechanisms have been invoked in the literature, such as the pyroelectric effect, Li diffusion or the thermoelectric effect.[29,39] In turn, due to the strong absorption, the penetration depth is low, leading to shallow domain structures. Conversely, femtosecond near-IR lasers work within the transparency window of LN, overcoming this limitation. In fact, they have recently unlocked the possibility to realize 3D ferroelectric domain structures in bulk LN.[42,43] This method relies on multiphoton absorption processes, hence requiring huge intensities achieved by tight spatiotemporal light focusing. However, this requisite restrains the scalability for large-area domain structures. In this case, domain inversion has been dominantly attributed to the thermoelectric effect, induced by steep temperature gradients at the laser focus.[41]

Herein, we report the first method for all-optical domain inversion by CW visible light irradiation in LN crystals. Most of this work is focused on iron-doped LN (Fe:LN), where Fe impurities introduce an absorption band in the visible spectrum. Despite this absorption band, Fe:LN crystals are partially transparent in the visible range at usual doping levels, allowing for deep penetration of light, similarly to methods based on near-IR light. Moreover, it is widely known that iron dopants boost the bulk PV effect of LN, leading to the highest saturation electric fields of all PV materials to date (up to ∼200 kV cm$^{-1}$).[46,47] While Fe:LN has been very well known and studied for many years in holography, only recently has it emerged as an outstanding optoelectronic platform for the light-driven actuation on different systems via its large electric fields. Indeed, the ability to generate customized electric field distributions by structured light, without external power supplies or lithography-patterned electrodes, makes Fe:LN a rather versatile and appealing asset. For example, Fe:LN has been successfully employed for the optofluidic manipulation of liquid droplets,[48–51] manipulation and patterning of micro/nanoparticles by PV optoelectronic tweezers,[52–56] guided locomotion and alignment of liquid crystals,[57,58] optical gating of graphene[59] or hybrid Fe:LN-graphene metasurfaces.[60,61] All these applications have been developed with monodomain Fe:LN crystals. However, multidomain structures could enhance the flexibility and potential of Fe:LN platforms, allowing the light-induced generation of arbitrary 2D bipolar charge distributions on *z*-cut substrates.

The method presented in this work offers an appealing scheme to easily tackle the flexible fabrication of such domain-engineered LN substrates. It consists in the light irradiation of the ferroelectric LN crystal while it is immersed in an electrically-conductive medium, such as water. Thus, our all-optical method stands out for its extreme simplicity and low cost: it only requires a CW visible light source, basic optics and a suitable liquid like water. These elements are ubiquitous, making it straightforwardly accessible for anyone. No power supplies or lithography-patterned electrodes are needed, as in conventional electric-field poling. Another advantage is that low/moderate optical power densities on the order of W cm$^{-2}$ are enough for all-optical domain inversion. Low-power visible lasers are easier to handle and align, safer and more affordable compared to UV, femtosecond near-IR or far-IR lasers. First, in Section 2 we provide a set of unequivocal tests that demonstrate all-optical domain inversion: pyroelectric trapping of charged microparticles, selective chemical etching and surface profilometry. In Section 3 we study the role of different parameters in the morphology of the inverted domains: exposure time, light intensity, focusing conditions and surrounding medium. Also, we characterize the 3D bulk structure of the domains by means of Čerenkov SHG microscopy. Remarkably, the deep penetration of light in the visible spectrum allows for domain structures deeper than any other all-optical technique without damaging the crystal (up to 289 ± 9 µm in this work). Finally, at the end of the manuscript (Section 4) we demonstrate that the method is not constrained to Fe:LN, but it can also be extended to undoped congruent LN.

## 2. Experimental Procedure for All-Optical Domain Inversion

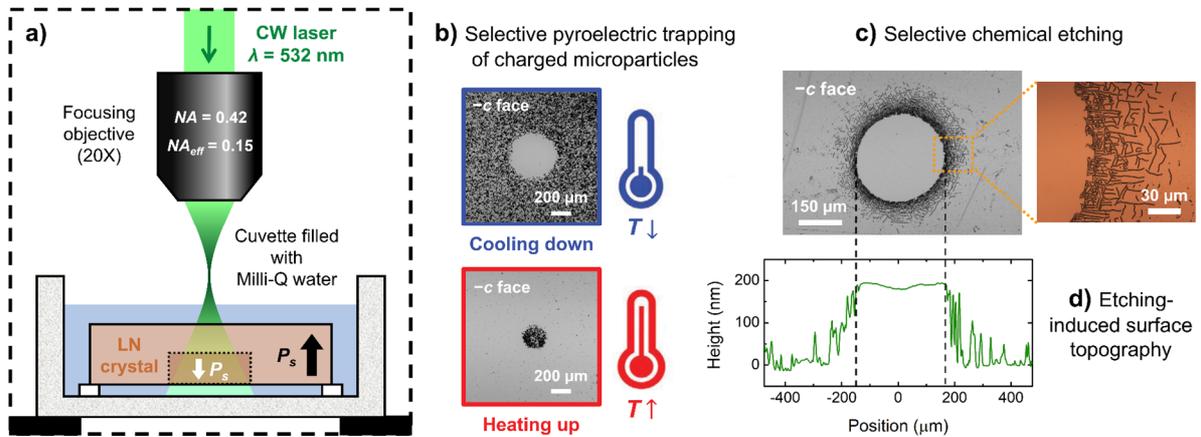

**Figure 1.** a) Simplified schematic of the experimental setup for all-optical domain inversion. b) Trapping/repulsion of positively-charged toner microparticles induced by the pyroelectric effect at the laser-irradiated area (intensity $I$ = 86.4 W cm$^{-2}$, exposure time $t$ = 1 min) of an Fe:LN crystal. The top image shows the result upon crystal cooling ($\Delta T$ = −12±1 °C), whereas the bottom image corresponds to crystal heating ($\Delta T$ = 12±1 °C). c) Optical bright-field image after selective chemical etching with HF acid. The inset shows the edge of the inverted domain with magnified detail. d) Measurement of the surface topography profile after selective chemical etching along the domain diameter.

## 2.1. Experimental Setup

As already mentioned, the experimental arrangement needed for all-optical domain inversion is strikingly simple. The basic setup employed in this work is illustrated in **Figure 1**a. Essentially, a CW laser operating at $\lambda$ = 532 nm is focused by an objective lens (magnification 20X and full numerical aperture $NA$ = 0.42). Since the input laser does not cover the full pupil of the objective, the effective numerical aperture is $NA_{eff}$ = 0.15. Then, the focused laser beam impinges on the LN sample, which is surrounded by ultrapure Milli-Q water inside a cuvette, except in Section 3.4 where different surrounding media are compared. The cuvette is mounted on top of an XYZ stage. Unless otherwise stated, the crystal is located below the beam waist, as illustrated in Figure 1a. In this configuration, the default spot size of the Gaussian beam has a $1/e^2$ diameter of 150 µm at the top surface and around 280 µm at the bottom, due to its divergence. This setup is embedded in a custom-made optical microscope (see Figure S1 in the Supporting Information).

In this work, we have predominantly used *z*-cut Fe:LN crystals, but also undoped congruent LN in Section 4. Both of them are initially monodomain and have a thickness of 1 mm. The Fe:LN crystals have a 0.25 mol% iron concentration and an absorption coefficient of $\alpha_{abs}$ = 20.4 cm$^{-1}$ at 532 nm. Prior to every experiment, the samples were carefully cleaned with distilled water and acetone, using lens-cleaning tissue. During the experiments, the crystals were always oriented so that the incident face of the incoming laser beam was the +*c* face, as shown in Figure 1a.

## 2.2. Demonstration of Light-Induced Domain Inversion

During light excitation of Fe:LN crystals surrounded by Milli-Q water, we observed in bright-field transmission microscopy that some material modifications were generated at the −*c* face, i.e. at the bottom non-incident face (see Video S1 in the Supporting Information). These modifications were stable and did not disappear over time after the experiment, even after the dark decay of the PV electric fields in a few days. Interestingly, nothing was observed at the +*c* face. Based on this preliminary hint, we hypothesized that ferroelectric domain inversion driven by light could be taking place. Thus, we carried out several tests.

First, we exploited the pyroelectric effect to decorate the ferroelectric domains using charged microparticles. Upon heating/cooling, the spontaneous polarization of LN decreases/increases, leading to net surface charges due to the imbalance between polarization and screening charges. Namely, negative/positive surface charges are induced at the +*c* face upon heating/cooling, whereas surface charges with opposite polarity are generated at the −c face.[14] This effect has been widely used for micro/nanoparticle patterning in the literature.[21] To make sure that there was no influence whatsoever from PV space charges after the light excitation, these experiments were conducted weeks later, much longer than the dark lifetime of the PV fields in our Fe:LN crystals. We used toner microparticles extracted from the cartridge of a laser printer (Canon, C-EXV 29, black toner). The size of the toner particles is around ∼1-10 µm and they carry positive net charge. Then, the toner particles were dispersed in an insulating liquid, namely *n*-heptane (PanReac AppliChem), using an ultrasonic bath. Prior to the pyroelectric trapping, the particle suspension was cooled down in a fridge ($\Delta T = -12 \pm 1$ °C) or heated up in a heating plate ($\Delta T = 12 \pm 1$ °C). The temperature of the suspension was measured with an IR thermometer and a thermocouple, yielding the same result. Finally, the LN crystal with inverted domains (initially at room temperature, $T = 21 \pm 1$ °C) was immersed in the cold/hot particle suspension for around 20 seconds. The results are shown in Figure 1b. Upon cooling, the positively-charged particles were repelled from the irradiated area and attracted everywhere else. However, the opposite behavior is observed upon heating. This finding strongly suggests that the −c face has been locally switched to +c by the laser. Also, the repulsion area upon cooling is larger than the trapping area upon heating, due to the additional electrostatic action of the non-inverted −c surface. This is the first solid evidence of all-optical domain inversion.

To unequivocally reveal the light-induced inverted domains, the LN crystals were subjected to selective chemical etching in HF acid (purchased from J.T.Baker, 49% concentration) for 15 min at room temperature. This standard technique is based on the differential etch rate of the polar faces in LN: +c faces are essentially unaffected, whereas −c faces are appreciably etched.[62] Both polar faces of the crystals were exposed to the acid during etching. A representative result is shown in Figure 1c, which clearly proves that domain inversion takes place at the −c face as a consequence of light excitation. Again, no inverted domains were observed at the +c face. Moreover, Figure 1c provides insights into the surface domain morphology: a large fully-inverted circle with an intricate halo of densely-packed "tentacles" at the edge, having typical widths in the range between ∼50 nm and ∼1 µm. High-resolution images of the domain edge may be found in Figure S3, S4 and S5 of the Supporting Information, taken with a Scanning Electron Microscope (SEM). Some tentacles have triangular protuberances, whose side can be up to ∼6 µm. Also, the orientation of the tentacle-shaped domains is not random: they point preferentially in the three crystallographic −y directions at different positions of the halo (see Figure S6).

Finally, we measured the surface topography profile after etching using a contact stylus profilometer (Dektak IIA). The height profile of Figure 1d indicates that the non-illuminated −$c$ surface is etched faster than the illuminated (inverted) spot, as expected. This result is again consistent with the fact that +$c$ domains are not appreciably affected by HF acid, whilst −$c$ domains are. Moreover, from the height of the step (189±18 nm), the differential etch rate can be estimated as 0.76±0.07 µm h$^{-1}$, in good agreement with the etching rate of the −$c$ face reported in the literature for undoped LN in similar conditions.[62] Therefore, the results of Figure 1 undoubtedly demonstrate all-optical domain inversion of Fe:LN crystals.

## 3. All-Optical Domain Inversion in Fe:LiNbO$_3$

### 3.1. Role of Exposure Time and Light Intensity

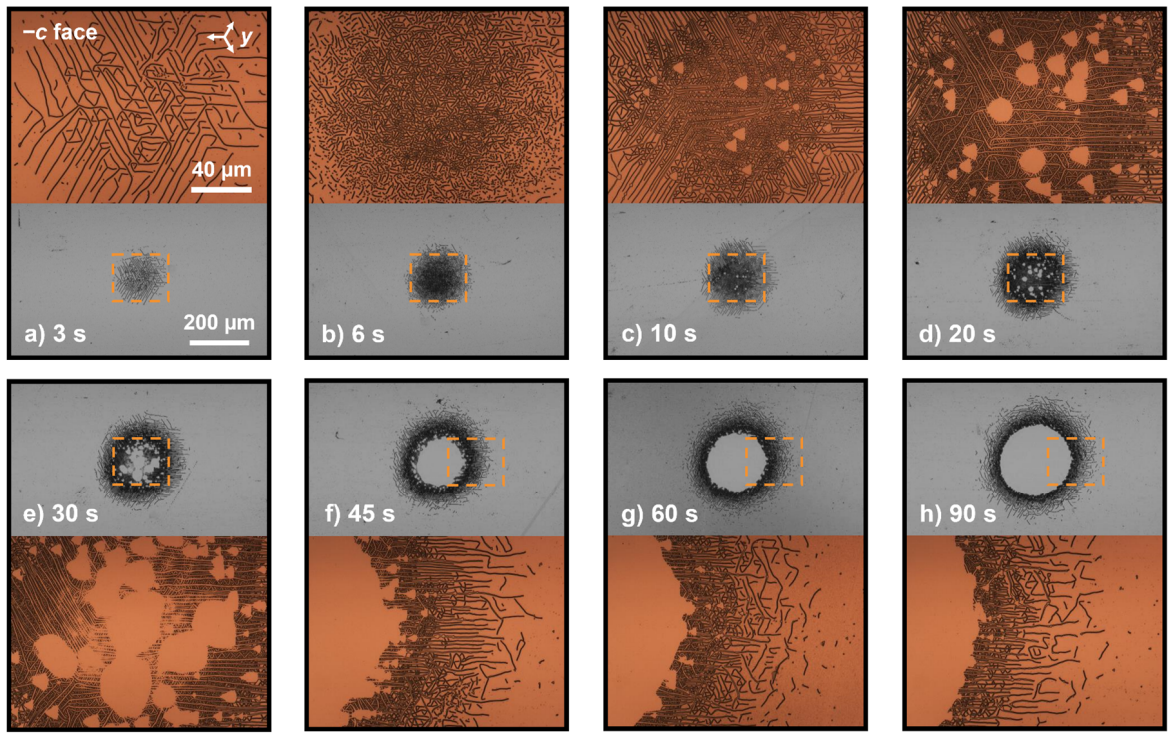

**Figure 2.** Optical bright-field images (after etching) of light-induced inverted domains for different exposure times from a) 3 s up to h) 90 s. In all cases the light intensity is $I = 28.8$ W cm$^{-2}$. The brown-colored images correspond to magnified areas of the widefield grayscale images (indicated by a dashed rectangle).

In this section, the influence of the illumination parameters in the process of ferroelectric domain inversion is studied. In **Figure 2**, eight experiments with different exposure times but fixed intensity are shown. Clearly, the morphology of the inverted domains strongly depends on the exposure time. Based on these results, one can establish the following kinetics of the domain formation during light excitation:

1) Nucleation starts in the form of sparse nanodomains in the illuminated area, as shown in Figure S4a for a lower intensity. Most of these submicron domains are irregular hexagons (quasi-triangles) with the vertices oriented in the *y* crystallographic directions, while some of them show irregular geometries.

2) Then, the nanodomains serve as seeds for the growth of self-assembled nanoscale maze-like structures (see Figure 2a). These intricate structures are not random, but preferentially grow in the *y* directions (see also 2D Fourier transform in Figure S7). The typical width of maze branches is in the range between ∼50 nm and about ∼1 µm. Also, the longer the exposure time, the thicker the branches and the more closely-packed the maze becomes (see Figure 2b). The emergence of such a complex domain structure is thought to be due to correlated nucleation in the irradiated area.[22]

3) When the density of domains in the maze is high enough, scattered triangular domains start to form (see Figure 2c), all of them oriented with the corners pointing in the +*y* crystallographic directions. The observed triangles have various sizes, and they grow over time (see Figure 2d). It is worth noting that triangular domains are commonly observed in LN due to the trigonal symmetry of the lattice. In some cases, one can see that the triangular shape is lost, evolving into rounded geometries. Eventually, coalescence between the triangles themselves takes place (see Figure 2e).

4) In the end, complete coalescence takes place at the center of the illuminated spot (see Figure 2f), producing a quasi-circular domain. Meanwhile, a halo of "tentacles" and triangular domains persists at the edge, inherited from the former maze. Over time, the inverted domain keeps growing sideways (see Figure 2g and 2h), increasing the diameter of the inner circle.

To shed more light on the role of the optical parameters, we conducted further experiments with different intensities and exposure times, shown in **Figure 3**. The first observation from these results is that the kinetics of domain formation are faster for increasing light intensities (see also Video S2). Furthermore, the intensities and some of the exposure times were deliberately chosen so that they were integer multiples of each other, with a multiplication factor of 3. As a result, one can compare spots with different light intensities and exposure times, but equal exposure $E = I \cdot t$. These cases are highlighted with frames of different colors in Figure 3a-3c, where one can qualitatively see that such spots are at the same stage of the domain inversion process in terms of morphology. This surprising feature is also demonstrated quantitatively in Figure 3d and 3e. When the diameter of the inner inverted circle is plotted as a function of exposure (Figure 3e) instead of exposure time (Figure 3d), all the curves merge into a single one. Notably, a time evolution governed by exposure is a well-known fingerprint of the bulk PV effect in the one-center model for Fe:LN.[63] Moreover, a threshold exposure of about ~620 J cm$^{-2}$ can be inferred from Figure 3e for the formation of quasi-circular domains due to coalescence.

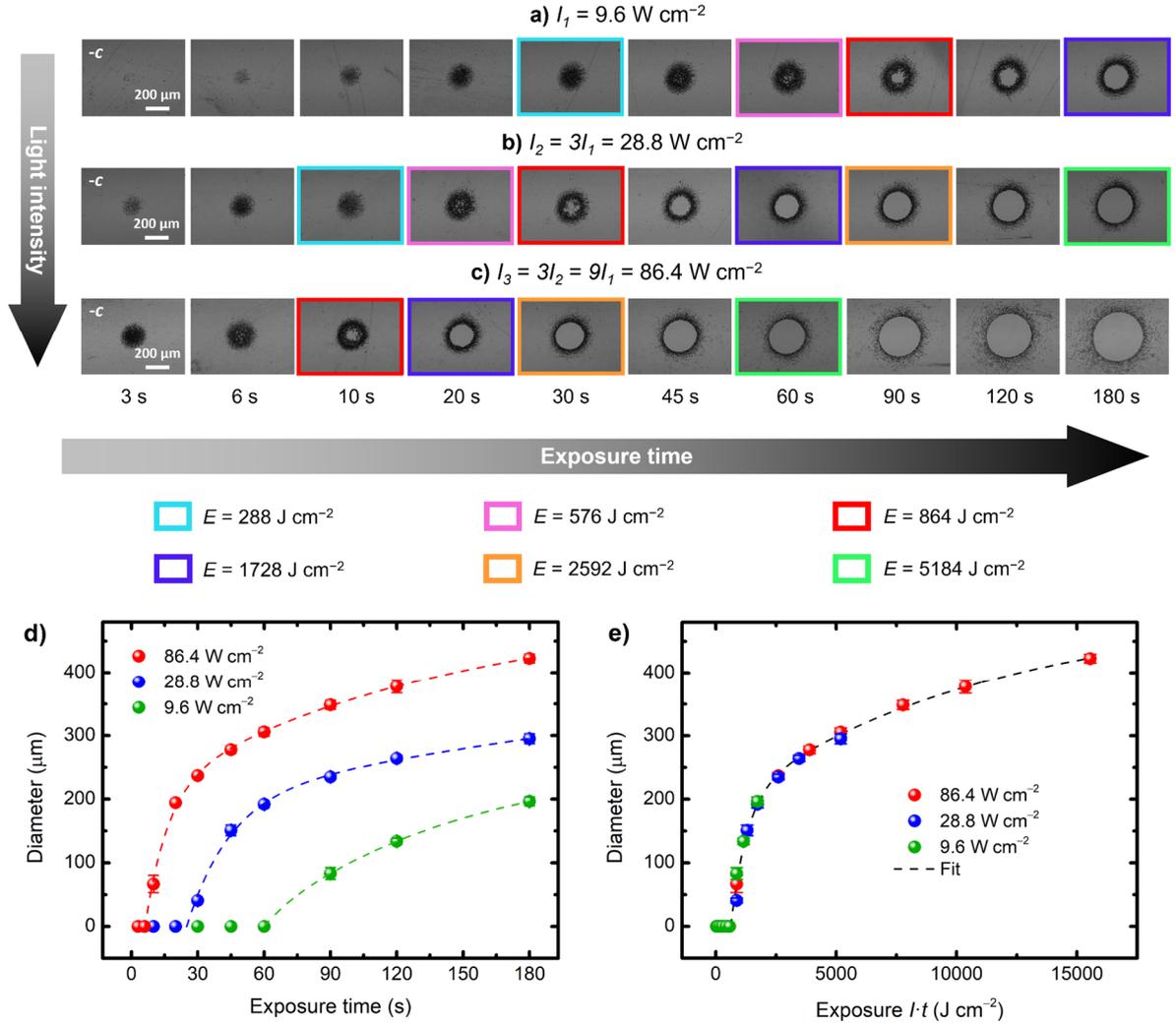

**Figure 3.** a)-c) Bright-field images of 30 domain inversion experiments using different light intensities and exposure times (after etching). Each row corresponds to a different light intensity ($I$) and each column to a different exposure time ($t$). The images with equally-colored frames correspond to results with identical exposures ($E = I \cdot t$). In the bottom graphs, the diameter of the inner circular domains is plotted as a function of d) exposure time and e) exposure. The dashed lines are fits (included as a visual guide) to a function $f(x) = A_1 \cdot (1 - e^{-(x-x_0)/x_1}) + A_2 \cdot (1 - e^{-(x-x_0)/x_2})$, where $x$ is either exposure time in d) or exposure in e).

### 3.2. 3D Characterization by Čerenkov SHG Microscopy

To gain insights into the depth of the light-inverted domains, we employed a Čerenkov SHG microscope, a widely-used technique for 3D mapping of ferroelectric domains.[64,65] Details about the setup and the measurements may be found in Section E of the Supporting Information.

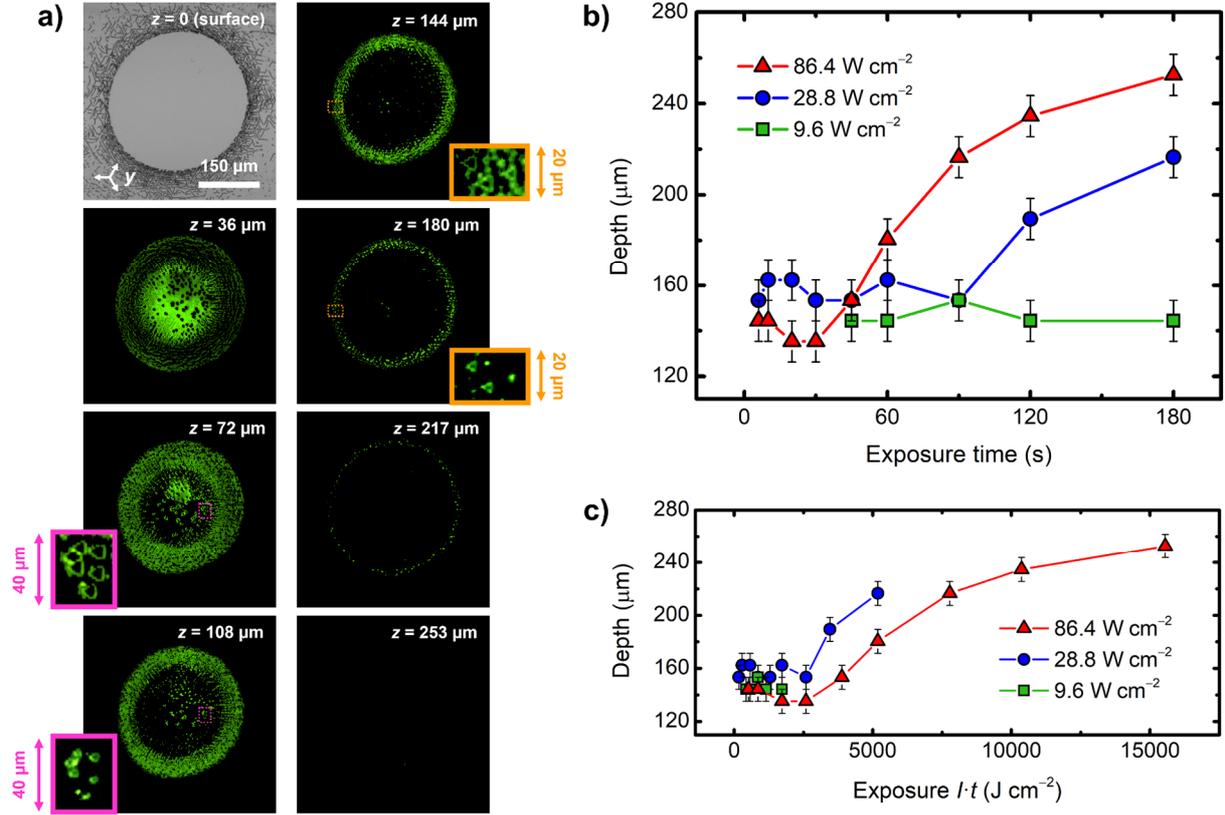

**Figure 4.** a) Representative measurements of one of the inverted spots of Figure 3 ($I$ = 86.4 W cm$^{-2}$, $t$ = 120 s) at different depths using a Čerenkov SHG microscope. The image for $z$ = 0 corresponds to etching at the surface. Green areas indicate the SHG signal coming from domain walls. The pink and orange insets are magnified areas from two different regions. The graphs show the maximum depth reached by the domains as a function of b) exposure time and c) exposure. The solid lines have been plotted as a guide for the eye.

The results of the Čerenkov SHG scans are shown in **Figure 4**. First of all, Figure 4a reveals a rather complex domain structure in the bulk of the Fe:LN crystal. On one hand, the "tentacles" around the inner circle are shallow, so we could not reliably measure the depth in our setup. The same applies to the nanoscale maze structures observed at low exposures (see Figure 2a). On the other hand, the inner circle may be subdivided in two parts: the central region and the outer ring. The central region comprises a large density of "spike" domains, even though it is uniformly inverted at the surface. The insets show the in-plane triangular shape of some of these domains, all of them oriented with the corners in the +$y$ directions as usual, consistent with the etching results. The size of these triangles diminishes with depth and can typically go as deep as ~150 µm. In contrast, the outer ring is only observed when exposure is long enough and it can go deeper than 150 µm (see Figure 4a). This ring is also an ensemble of densely-packed spike domains, having in-plane triangular shape. The deeper into the crystal volume, the thinner the ring gets until it vanishes. In these illumination conditions, this ring has reached a maximum depth of 253±9 µm for the highest intensity and longest exposure time, as shown in Figure 4b. We believe that this ring of enhanced depth could be related to the accumulation of PV charge at the edge of a Gaussian beam, revealed by numerical simulations under ideal open-circuit conditions in ref. [66].

In Figure 4b and 4c, the maximum depth of the domains has been plotted as a function of exposure time and exposure, respectively. Overall, the general trend of the axial growth along the $z$ direction is somewhat similar to the surface lateral growth studied in Section 3.2. However, an interesting property is that the maximum depth is not solely governed by exposure alone ($E = I \cdot t$), unlike the diameters of Figure 3. In other words, intensity and time do not play an interchangeable role when it comes to the domain depth, since not all experimental data are superimposed in Figure 4c. This result suggests that, aside from the bulk PV effect in the one-center regime, there are additional contributions involved in the axial growth of the light-inverted domains along the polar direction $z$. There are two possible contributions: a second PV center and the pyroelectric effect. First of all, it is known that above intensities of ~1000 W cm$^{-2}$, the saturation PV field becomes intensity-dependent due to Nb$_{Li}$ antisites, which act as a second PV center apart from Fe impurities.[46,63,67] As a result, the time evolution is no longer determined by exposure alone. In our experiments this threshold intensity is not reached at the $-c$ face, but it is indeed surpassed at the incident $+c$ face up to a certain depth into the crystal volume, due to a smaller spot size caused by the beam divergence and a lower absorbance. Also, in this intensity regime, the Fe:LN crystal could be locally heated up due to absorption, leading to pyroelectric effect. Anyhow, further work is necessary to clarify this matter.

### 3.3. Influence of the Laser Spot Size

The impact of tighter light focusing conditions was also explored. For that purpose, the waist of the laser beam was focused on the $-c$ face of the crystal, as shown in Figure S9a. Since the effective numerical aperture in our setup is $NA_{\text{eff}} = 0.15$, the 1/e$^2$ diameter of the diffraction-limited light spot is estimated as $d = 2\lambda/\pi NA_{\text{eff}} = 2.3$ µm. The Rayleigh length can also be estimated as $z_R = n_o \lambda / \pi NA_{\text{eff}}^2 = 17$ µm, where $n_o = 2.32$ is the ordinary refractive index of LN at 532 nm. In this configuration, domain inversion was only observed at the −c face (the bottom face), consistent with the behavior so far. The etching results are shown in **Figure 5** for different light intensities and exposure times.

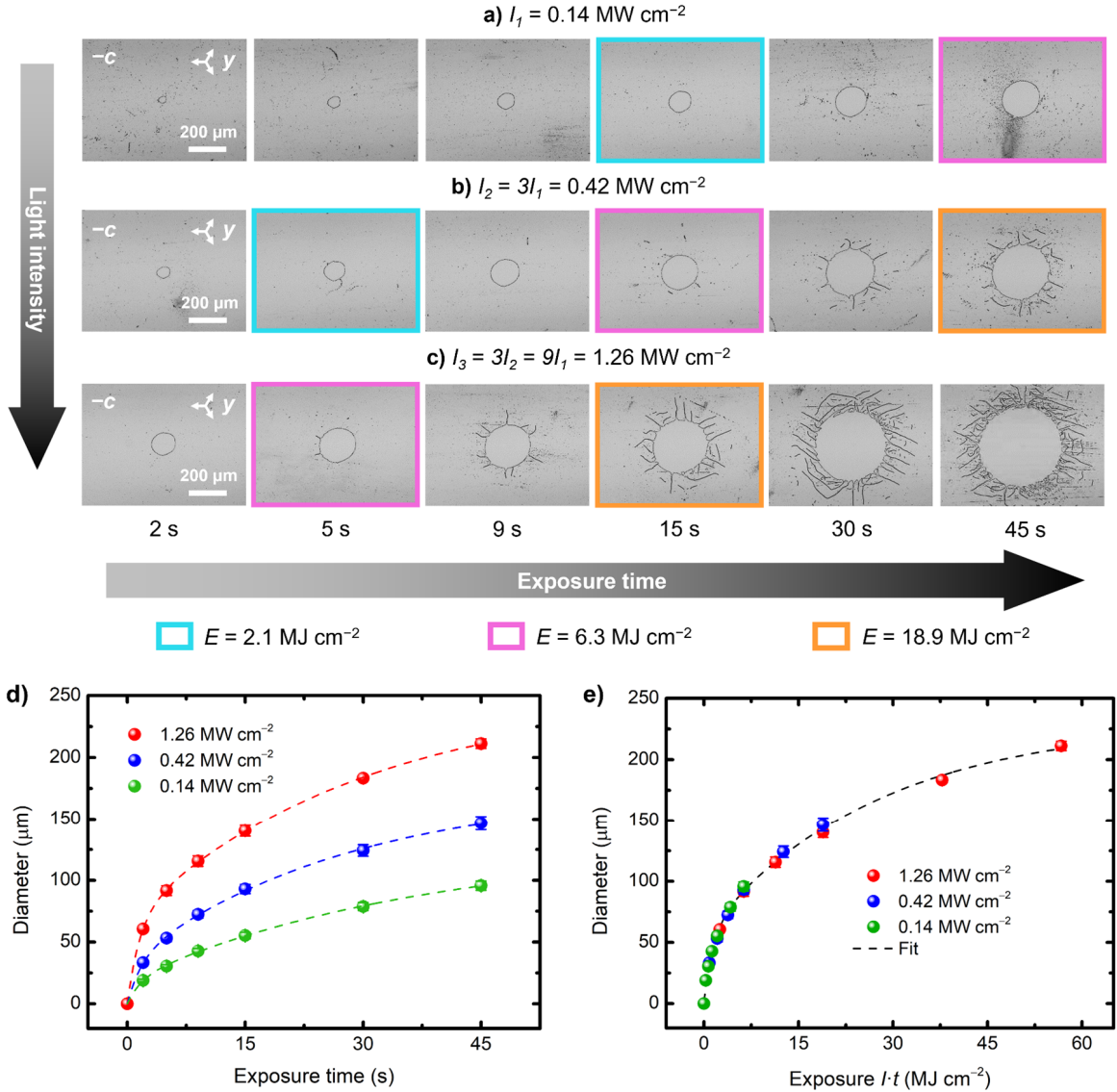

**Figure 5.** a)-c) Bright-field images of 18 domain inversion experiments using different light intensities and exposure times (after etching). In this case, the waist of the focused laser beam was located at the −c face, as illustrated in Figure S9a. Each row corresponds to a different light intensity (*I*) and each column to a different exposure time (*t*). The images with equally-colored frames correspond to results with identical exposures ($E = I \cdot t$). In the bottom graphs, the diameter of the inner circular domains is plotted as a function of d) exposure time and e) exposure. The dashed lines are fits (included as a visual guide) to a function $f(x) = A_1 \cdot \left(1 - e^{-(x-x_0)/x_1}\right) + A_2 \cdot \left(1 - e^{-(x-x_0)/x_2}\right)$, where *x* is either exposure time in d) or exposure in e).

In this case, the morphology of the optically-inverted domains differs from that of Figure 2 with a larger weakly-focused laser spot. For low exposures, quasi-circular domains are formed, with growing diameters over time (see Figure 5a and 5b). The domain wall is not perfectly rounded but full of small steps generated during the lateral growth (see Figure S9b). From a practical point of view, the ability to create clean quasi-circular domains, unencumbered by adjacent haloes, provides valuable added flexibility. In contrast, for long exposures, a halo of "tentacles" begins to appear at the border of the circle, somewhat similar to those of Figure 2. The number and length of tentacles increases with exposure time. Therefore, light focusing conditions provide another route to control the shape of inverted domains.

Another significant observation is that, even in this regime of high intensities (~MW cm$^{-2}$) several orders of magnitude higher than in Figure 3, the time evolution of the domain diameters is still governed by exposure ($E = I \cdot t$), both qualitatively and quantitatively, as evidenced in Figure 5d and 5e. Thus, intensity and exposure time play an equivalent role, a common feature of the bulk PV effect. It is also worth noting that in Figure 5d-5e, unlike Figure 3d-3e, no threshold for the formation of quasi-circular domains is visible. This is because at these high intensities this threshold exposure is reached in a very short timescale.

For further characterization, we also made additional Čerenkov SHG measurements, illustrated in **Figure 6**. In Figure 6a, the bulk domain structure is somewhat different with respect to Figure 4. In this case, the in-plane geometry of the "spike" domains is not triangular but irregular, and mostly elongated along *y* directions. However, all other relevant features are analogous. Namely, the outer part of the circular domain can grow deeper along the *c*-axis than the central part, giving rise to a ring of domains at the edge that gets thinner along *z* until it disappears. In this configuration with a focused laser beam, we have measured a remarkable maximum depth of 289 ± 9 µm (see Figure 6b). Moreover, even deeper domains can probably be attained by using higher intensities and longer exposure times. On the other hand, analogously to Figure 4, we have found that exposure alone does not determine the maximum depth of the domains (see Figure 6c), unlike the lateral growth at the surface (see Figure 5).

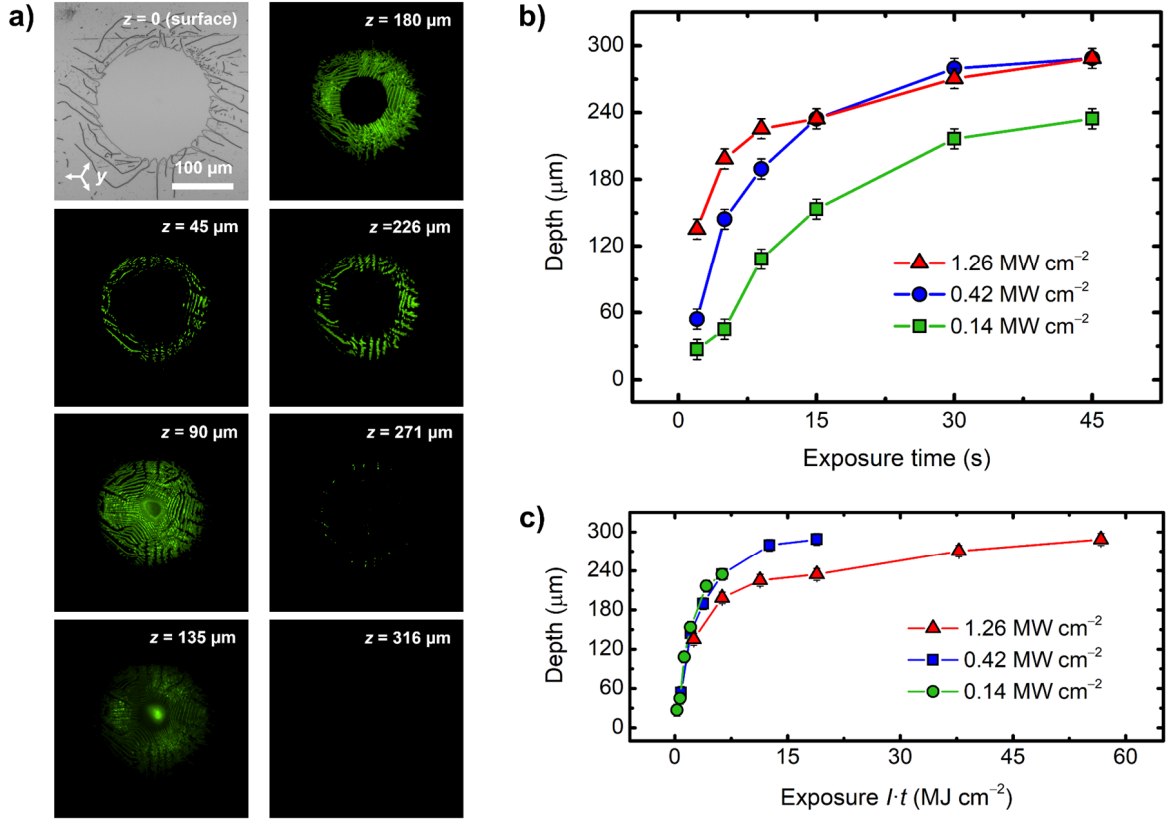

**Figure 6.** a) Representative measurements of one of the inverted spots of Figure 5 ($I = 1.26$ MW cm$^{-2}$, $t = 30$ s) at different depths using a Čerenkov SHG microscope. The image for $z = 0$ corresponds to etching at the surface. Green areas indicate the SHG signal coming from domain walls. The graphs show the maximum depth reached by the domains as a function of b) exposure time and c) exposure. The solid lines have been plotted as a guide for the eye.

### 3.4. Discussion on the Role of the Surrounding Medium

So far, all the results presented in the paper were obtained with Milli-Q water surrounding the Fe:LN crystal during the irradiation process. In this section we compare different surrounding media (see **Figure 7**). First of all, equivalent results were obtained when using Milli-Q water, tap water, acetone (Uvasol, purity ≥ 99.9%) or ethanol (PanReac AppliChem, purity ≥ 99.9%), as shown in Figure 7a and Video S3. More specifically, there is a uniform inverted circle in the middle with a "halo" of closely-packed domains at the border, in agreement with the behavior described in previous sections. These four liquids have in common a high electrical conductivity ($\gtrsim 10^{-5}$ S m$^{-1}$) and, therefore, fast electrostatic screening. Conversely, a radically different behavior was observed when using highly insulating liquids with low electrical conductivity, such as n-heptane (PanReac AppliChem, purity ≥ 99.0%) or liquid paraffin oil (PanReac AppliChem, pure), as shown in Figure 7b. In this case, only a residual amount of new micro/nanodomains appeared at the irradiated region, having varied irregular shapes and located at seemingly random locations. Moreover, marginal domain inversion was also noticed at the +$c$ face. Likewise, air (with a relative humidity of 28 ± 1 %) shows a similar behavior, but a larger number of triangular microdomains is induced at the light spot (see Figure 7c).

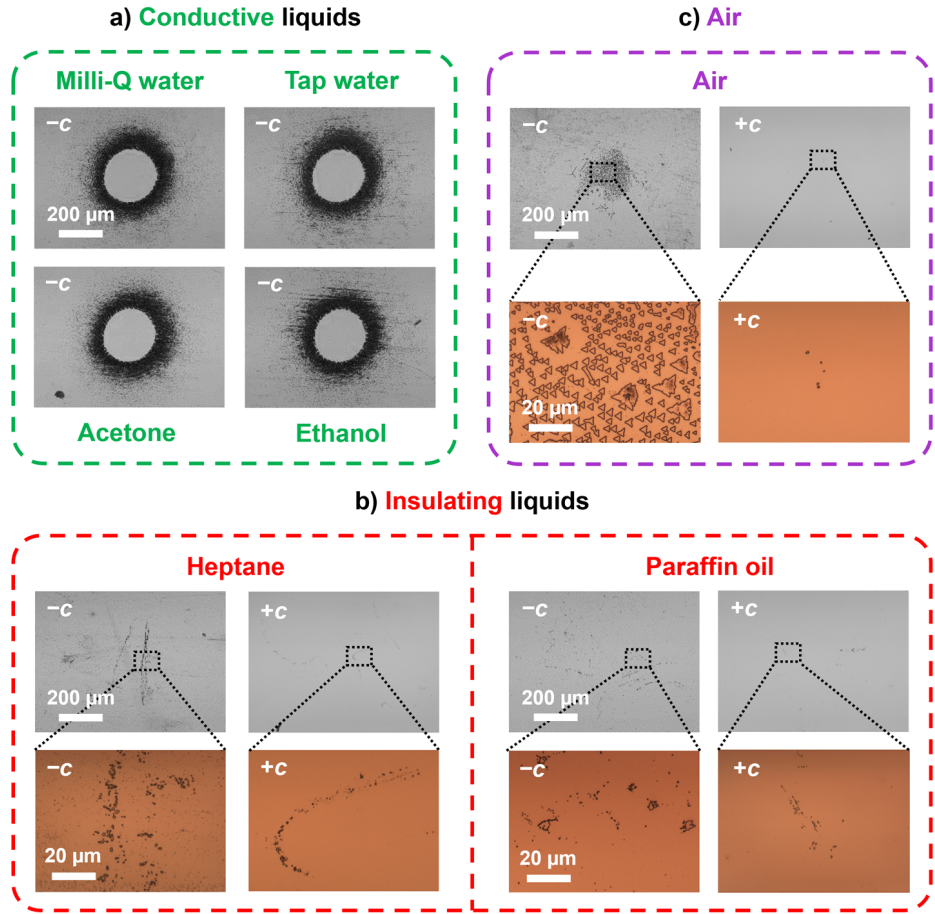

**Figure 7**. Bright-field images (after etching) of several experiments in different surrounding media: a) conductive liquids, b) insulating liquids, and c) air. In all cases, the exposure time is 3 min, the incident optical power is 81 mW and the light spot is at the center of the images (with a diameter of $d = 280$ µm at the −$c$ face, as explained in Section 2.1). The brown-colored images correspond to magnified areas of the widefield grayscale images.

Hence, these experiments reveal the critical role of the surrounding medium in the inversion process. We attribute this behavior to the electrical conductivity of the medium, which controls the external screening of the light-induced electric fields in the Fe:LN ferroelectric crystal. In fact, the residual amount of domain inversion observed in low-conductivity media could be due to dielectric breakdowns, which would make the medium conductive for a short amount of time until the evanescent electric field is rapidly screened. There are two main sources of light-induced electric fields: the bulk PV effect and the pyroelectric effect (due to absorption). Experimental evidence suggests that the bulk PV effect is the main driving agent because:

1) Domain inversion can be achieved even at low light intensities, when the crystal heating is expected to be negligible. So far, we have observed domain inversion with intensities as low as ~200 mW cm$^{-2}$ (incident power of ~1 mW) at the expense of increasing the exposure time accordingly ($t$ = 5 min or longer). This is attributed to the continuous supply of PV current under light excitation, unlike the pyroelectric effect which is not relevant at such intensities.

2) Domains can also be inverted in undoped congruent LN (explained in Section 4), where the absorption of a CW visible laser at the intensities of this work is negligible for heating purposes.

3) The lateral growth of the inverted domains is governed by exposure ($E = I \cdot t$), as shown in Figure 3 and 5.

Nonetheless, we cannot rule out secondary contributions from the pyroelectric effect at high intensities. For example, it could be related to our observation that depth is not solely described by exposure (Figure 4 and 6). However, both PV and pyroelectric fields upon heating are parallel to the spontaneous polarization in open-circuit conditions. PV fields under strong screening, however, remain poorly understood, and this is exactly the key element for all-optical domain inversion. Here, it is expected that the crystal-liquid screening interface plays a crucial role in the inversion process. Anyhow, further work is necessary to unravel the complete physical mechanism, which is out of the scope of this first report on the subject.

## 4. Outlook: All-Optical Domain Inversion in Undoped Congruent LiNbO$_3$

Finally, we explored the possibility of performing all-optical domain inversion in undoped congruent LN crystals. First, we used the same illumination conditions as in Figure 1, with a spot diameter of about 280 µm at the bottom face (the −$c$ face). Using the maximum output power of the laser, a maximum intensity of ~1.5 kW cm$^{-2}$ could be reached with this spot size. However, no domain inversion whatsoever was observed for exposure times up to 45 s, i.e. exposures up to 68 kJ cm$^{-2}$. Note that this is a key difference compared to Fe:LN crystals, where much lower intensities can easily induce domain inversion.

Then, due to the limitation of the laser output power, we progressively decreased the laser spot size in order to further increase the light intensity. By doing so, we were indeed able to locally invert domains, as illustrated in **Figure 8**. For example, for an intensity of 5.9 kW cm$^{-2}$, we observed a self-assembled maze structure (see Figure 8a), rather similar to the results with Fe:LN in the low-exposure regime (see Figure 2a). In this structure, the *y* directions are somewhat favored. As the intensity increases, the branches of the maze get thicker at the center of the irradiated area (see Figure 8b). Then, coalescence between the branches starts to take place (see Figure 8c). When the intensity is high enough, a uniformly-inverted quasi-circular domain is generated at the center, with a halo of tentacles preferentially pointing in the −*y* directions (see Figure 8d). In general, these structures are quite similar to Fe:LN in Figure 3 and Figure 5, and they are only generated at the −*c* face. Thus, the general trend is similar to Fe:LN.

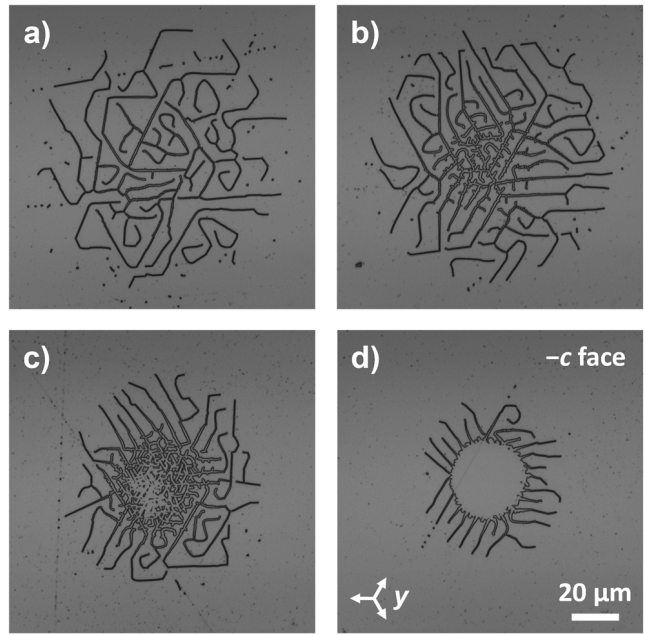

**Figure 8.** Bright-field images after etching of different domain inversion experiments in undoped congruent LN. Each experiment corresponds to a different laser spot size but equal incident optical power: a) *d* = 94 µm, *I* = 5.9 kW cm$^{-2}$, b) *d* = 84 µm, *I* = 7.4 kW cm$^{-2}$, c) *d* = 60 µm, *I* = 14.5 kW cm$^{-2}$, d) *d* = 36 µm, *I* = 40.3 kW cm$^{-2}$. In all cases the exposure time is 45 s.

Therefore, this technique is not just restricted to Fe:LN, but it can be generalized to undoped congruent LN as well. At this point, it is worth noting that the linear absorption is vanishingly low in undoped LN. However, the presence of intrinsic and extrinsic defects in LN leads to a residual absorption in the visible spectrum. In particular, trace impurities like Fe or Cu (on the order of ∼ppm) and intrinsic defects (namely $Nb_{Li}$ antisites) are known to be responsible for the bulk PV effect in nominally undoped LN.[63,68] Comparatively, PV currents are much lower than in Fe:LN. On the other hand, the contribution of the antisites leads to a saturation PV field that grows with light intensity, unlike Fe:LN where the saturation field remains constant in the one-center regime (i.e. low intensities). At high light intensities, PV fields in undoped congruent LN become close to those of Fe:LN, reaching values up to ∼200 kV cm$^{-1}$ at ∼$10^6$-$10^7$ W cm$^{-2}$.[46] Thus, the inherently lower PV currents in undoped LN and the intensity-dependent PV field could account for the higher intensities needed to invert domains compared to Fe:LN. A more detailed experimental analysis of undoped congruent LN will be tackled in future work, including the depth and the role of exposure time.

## 5. Conclusions

Overall, this paper encompasses a set of solid proofs that demonstrate for the first time all-optical ferroelectric domain inversion of Fe:LN and undoped LN crystals with visible light. According to our results, the successful realization of our method strongly relies on the presence of a conductive surrounding medium during irradiation. Although the underlying physical mechanism remains unclear, it is mainly attributed to the interplay between the bulk PV effect and external electrostatic screening from the surrounding medium. At this stage, additional contributions from the pyroelectric effect cannot be ruled out. Further research to unravel the physics of the process is currently ongoing. Anyhow, the experimental results show that the morphology of the inverted domains can be flexibly controlled by the light intensity, exposure time or focusing conditions, all the way from self-assembled maze structures to quasi-circular domains. Notably, domains with a maximum depth of 289 ± 9 µm have been measured, deeper than any other IR/UV all-optical technique to date without crystal damage.

The method is exceptionally simple, versatile and easy to implement, not requiring any lithography-patterned electrodes or external voltage supplies. Moreover, since the bulk PV effect of LN can be excited in the full visible spectrum, there is a broadband gamut of suitable light sources compatible with the method. In fact, incoherent light sources with moderate optical powers could also be employed. Likewise, spatial light modulators could be readily included in the setup to tailor complex light patterns, aiming at the parallel fabrication of arbitrary ferroelectric domain structures.

These domain structures are of practical interest for a number of fields, such as nonlinear photonics. Also, optically-patterned domains could be used as templates for mask-free lithography of Fe:LN and LN surfaces via selective wet etching. For instance, ridge waveguides[29] or LN metasurfaces[8] may be envisioned. Likewise, PV optoelectronic tweezers based on Fe:LN could strongly benefit from a completely new degree of freedom not explored so far: multidomain Fe:LN substrates. This feature could enhance already existing applications or pave the way towards new functionalities of PV platforms.

**Acknowledgements**

Financial support for this research project is gratefully acknowledged: grant PID2020-116192RB-I00 by MCIN/AEI/10.13039/501100011033 and grant TED2021-129937B-I00 by MCIN/AEI/10.13039/501100011033 and EU (FEDER, FSE). C. Sebastián-Vicente also thanks financial support through his FPU contract (FPU19/03940) and his grant for a research stay in Münster (EST23/00644). Also, we acknowledge the service from the MiNa Laboratory at IMN, and funding from CM (project S2018/NMT-4291 TEC2SPACE), MINECO (project CSIC13-4E-1794) and EU (FEDER, FSE). Likewise, we would like to express our gratitude to Fernando Moreno for kindly conducting the surface profilometry measurements.

Supporting Information

**All-Optical Domain Inversion in LiNbO₃ Crystals by Visible Continuous-Wave Laser Irradiation**

*Carlos Sebastián-Vicente, Jörg Imbrock, Simon Laubrock, Olga Caballero-Calero, Angel García-Cabañes, and Mercedes Carrascosa\**

### A. Full Optical Setup and Further Experimental Details

A schematic diagram of the experimental setup employed for all-optical domain inversion is shown in Figure S1. Note that the LN crystal is actually immersed in a cuvette as shown in Figure 1, typically filled with Milli-Q water unless otherwise specified.

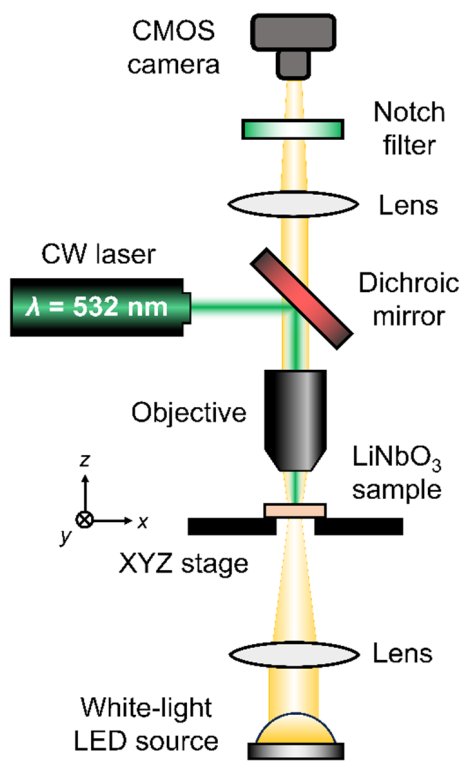

**Figure S1**. Complete experimental setup for all-optical domain inversion.

Regarding light intensities, they are always specified at the bottom face of the crystal, i.e. the $-c$ face where domain inversion takes place in our experimental conditions. This means that losses due to Fresnel reflections and absorption along the crystal thickness should be taken into account. Intensities are calculated with the following expression:

$$I = \frac{P \cdot (1 - R_{aw})(1 - R_{wc})e^{-\alpha_{abs}h}}{\pi(2\sigma)^2} \quad (S1)$$

where $P$ is the incident optical power (measured with a thermopile), $R_{aw}$ is the Fresnel reflectance at the air-water interface, $R_{wc}$ is the Fresnel reflectance at the water-crystal interface, $\alpha_{abs}$ is the absorption coefficient of the LN crystal at 532 nm, $h$ is the crystal thickness and $2\sigma$ is the $1/e^2$ radius of the Gaussian beam. Secondary Fresnel reflections are neglected. It is worth noting that the peak intensity of the Gaussian beam would be $2I$. The coefficients $R_{aw}$ and $R_{wc}$ can be computed using the Fresnel formula at normal incidence: $R = [(n_1-n_2)/(n_1+n_2)]^2$, where $n_1$ and $n_2$ are the refractive indices of the media forming the interface. This expression yields $R_{aw} = 0.02$ and $R_{wc} = 0.07$. On the other hand, the absorption correction factor in our crystals is $e^{-\alpha_{abs}h} = 0.13$. Thus, for instance, for an incident power of 50 mW and a typical laser spot size of $d = 4\sigma = 280$ µm, the intensity would be $I = 9.6$ W cm$^{-2}$.

## B. Supporting Videos

**Video S1.** Real-time video of the irradiation of a Fe:LN crystal ($I = 28.8$ W cm$^{-2}$, $t = 180$ s). The speed of the video has been increased by a factor 12X.

**Video S2.** Real-time video of the irradiation of a Fe:LN crystal for several light intensities: a) $I = 9.6$ W cm$^{-2}$, b) $I = 28.8$ W cm$^{-2}$, and c) $I = 86.4$ W cm$^{-2}$. In all cases the exposure time was $t = 90$ s. The speed of the video has been increased by a factor 6X.

**Video S3.** Real-time video of the irradiation of a Fe:LN crystal surrounded by several liquid media with high electrical conductivity: a) Milli-Q water, b) tap water, c) acetone, and d) ethanol. In all cases the exposure time was $t = 180$ s and the incident optical power was 81 mW. The speed of the video has been increased by a factor 12X.

In Videos S1-S3 (transmission bright-field microscopy) the domain inversion process can be observed in real time at the non-incident face of the crystal (i.e. the −$c$ face). Snapshots of Video S1 are given in Figure S2 as an example. Note, however, that the microscope image is focused at the upper incident face (i.e. the +$c$ face). As a result, the optical image is not sharp at the face where light-induced domain inversion occurs.

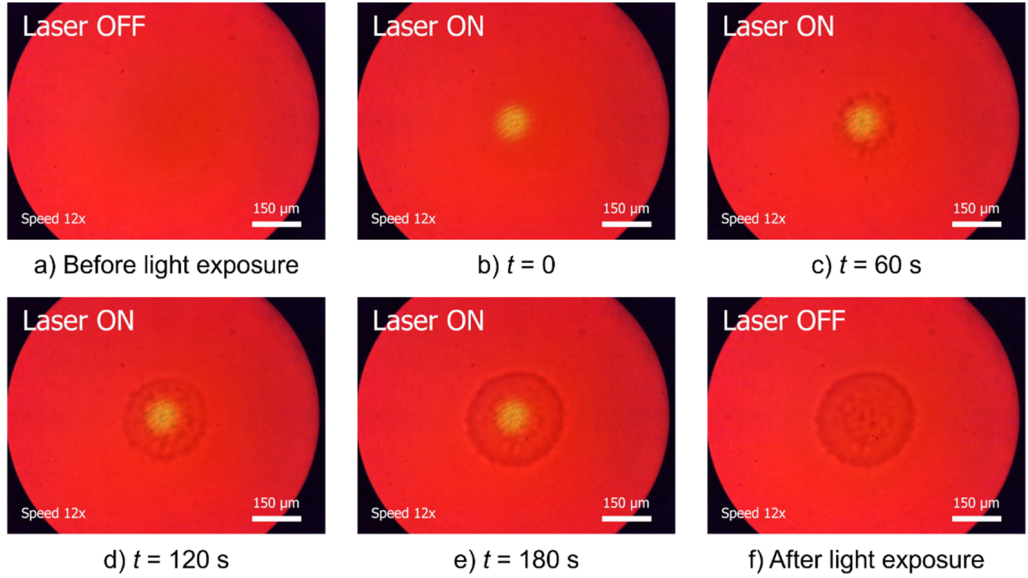

**Figure S2.** Sequence of snapshots corresponding to Video S1.

## C. Scanning Electron Microscopy (SEM)

Due to the nanoscale feature sizes of some of the domain structures, high-resolution SEM images of the etched domains were acquired. Prior to coating the sample, the crystals were gently cleaned using lens-cleaning tissue soaked in distilled water mixed with soap (Fairy). The sample was finally rinsed with bare distilled water and the liquid residuals were blown away with compressed air. Then, a 10-nm-thick conductive coating of Chromium was deposited on the −$c$ face of the crystal by sputtering, to prevent charge accumulation during SEM inspection. Finally, a high-resolution SEM was employed (model Verios 460, from FEI).

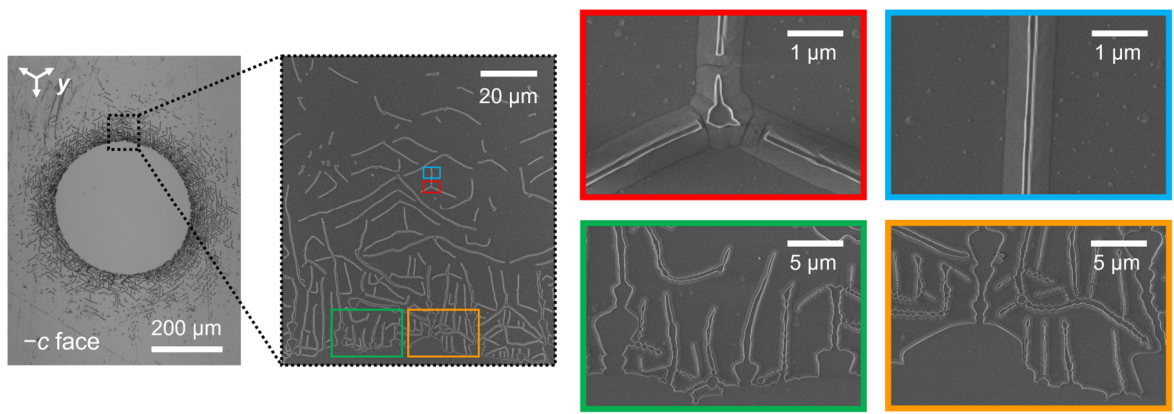

**Figure S3.** High-resolution details of the tentacle domains at the edge of a circular domain (intensity $I = 86.4$ W cm$^{-2}$, exposure time $t = 2$ min). The image on the left corresponds to the bright-field image after etching, while the others are SEM images at different magnifications.

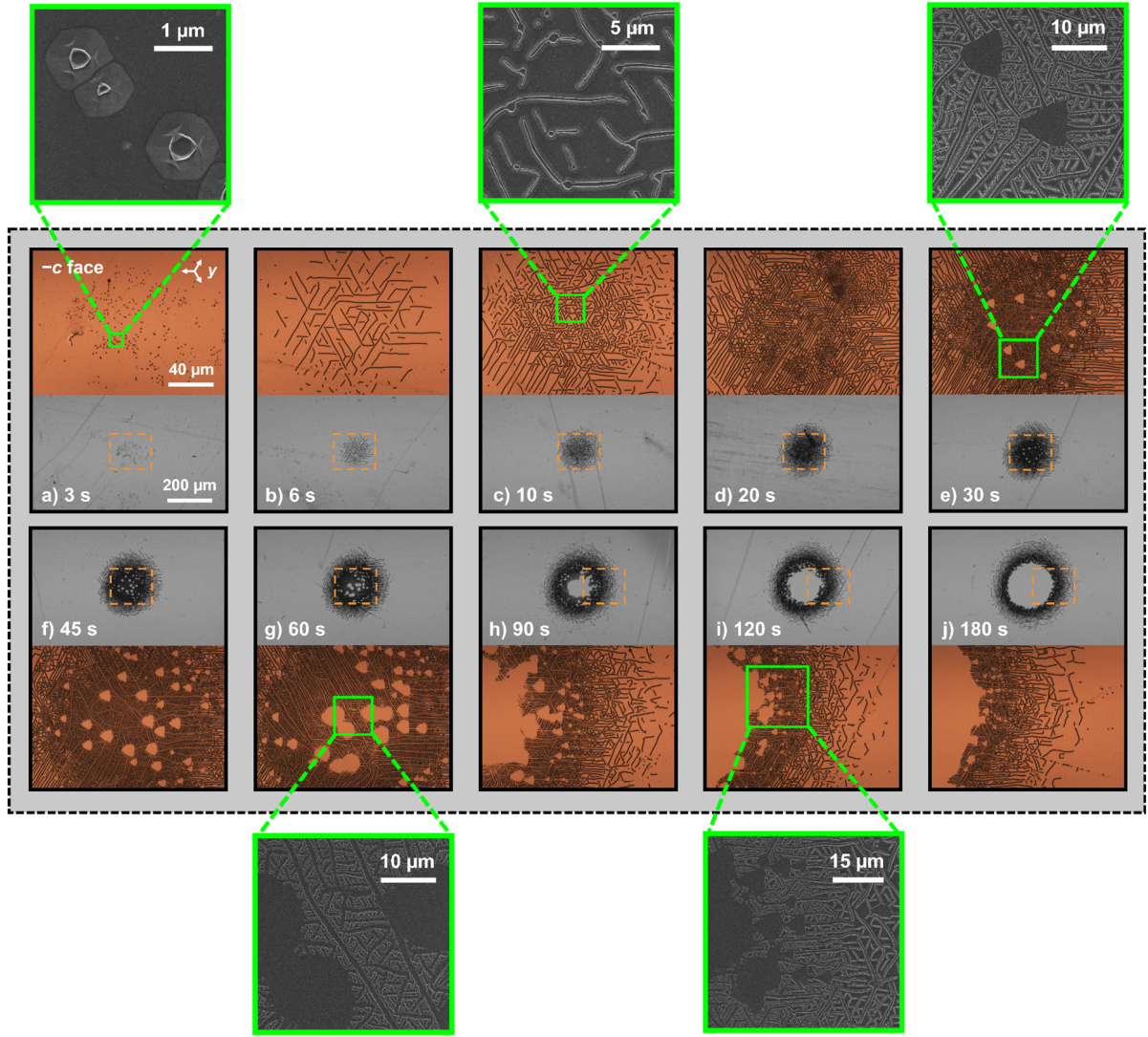

**Figure S4.** Micro-photographs (after etching) of light-induced inverted domains for different exposure times from a) 3 s up to h) 180 s. In all cases the light intensity is $I$ = 9.6 W cm$^{-2}$. The brown-colored images correspond to magnified areas of the widefield grayscale images (indicated by a brown dashed rectangle). The green insets correspond to high-resolution SEM images.

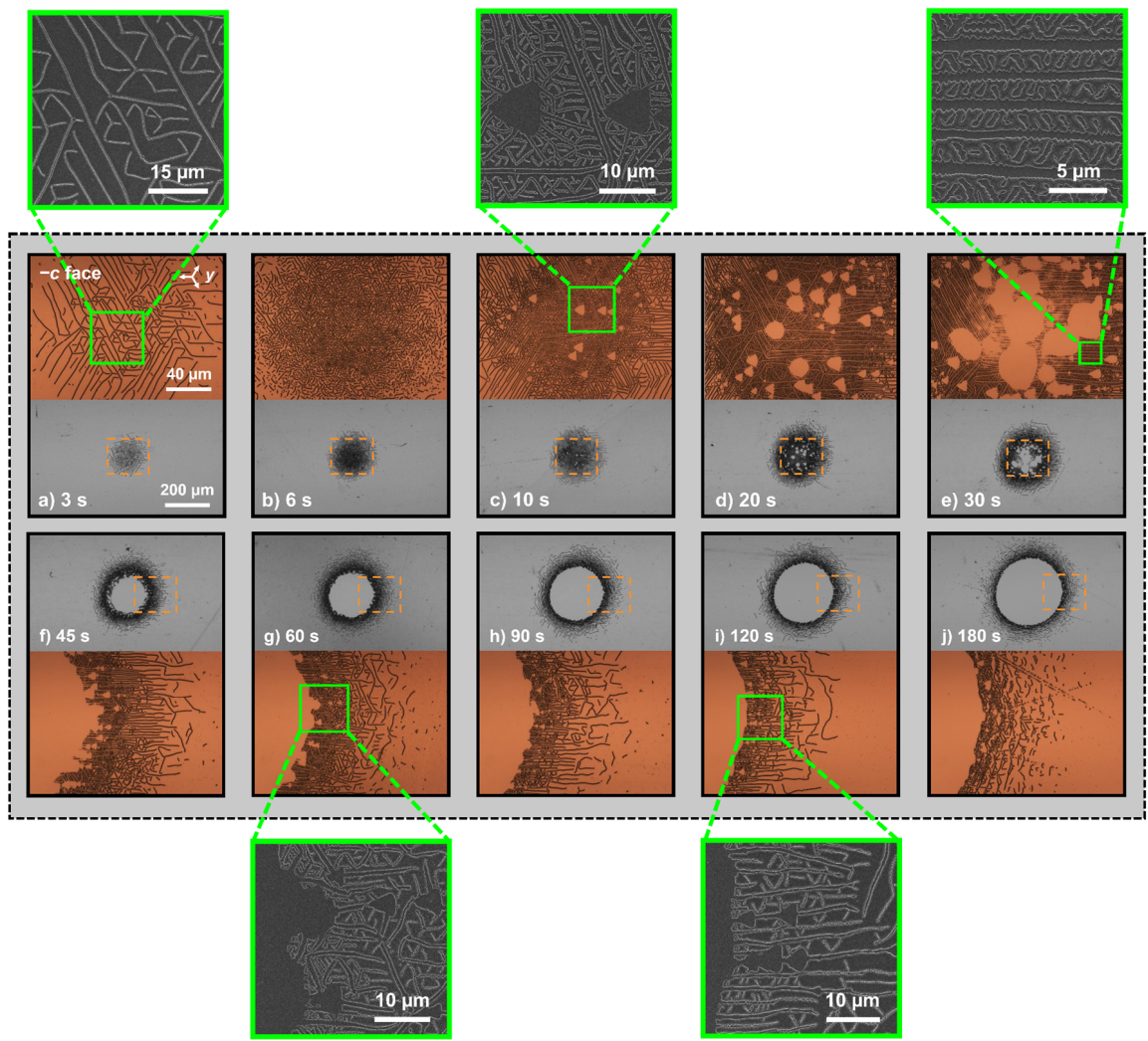

**Figure S5.** Micro-photographs (after etching) of light-induced inverted domains for different exposure times from a) 3 s up to h) 180 s. In all cases the light intensity is $I = 28.8$ W cm$^{-2}$. The brown-colored images correspond to magnified areas of the widefield grayscale images (indicated by a brown dashed rectangle). The green insets correspond to high-resolution SEM images.

## D. Orientation of "Tentacle" Domains and Self-Assembled Maze Structures

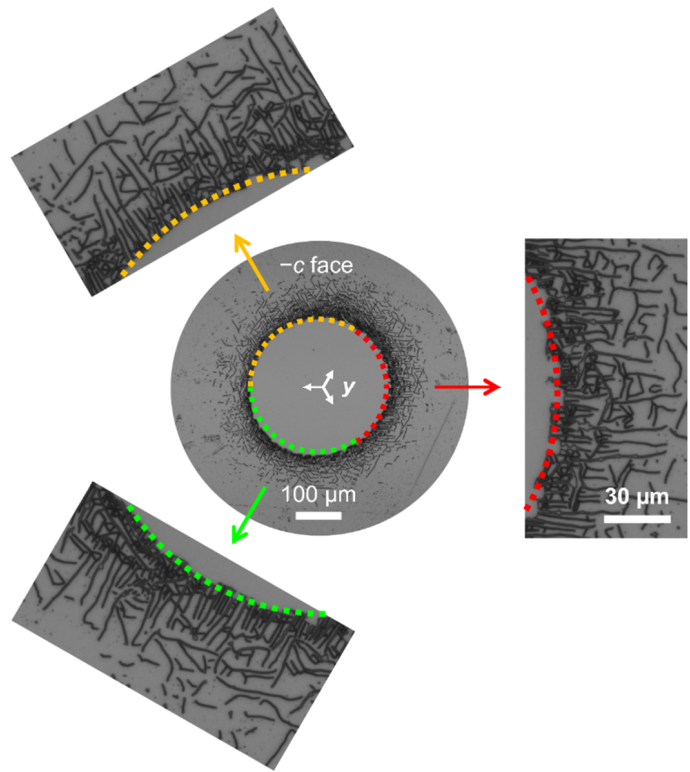

**Figure S6.** Illustration of the directionality of the "tentacle" domains located at the border of a circular domain ($I = 86.4$ W cm$^{-2}$, exposure time $t = 1$ min). Each sector of the halo with a given orientation is highlighted by a different color. There are three main orientations, coinciding with the $-y$ crystallographic directions.

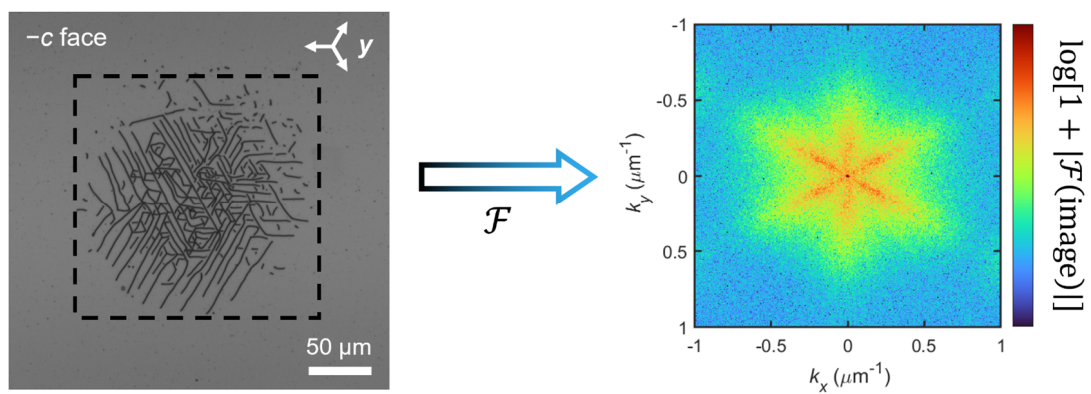

**Figure S7.** Example of a 2D Fourier transform (right) of a self-assembled maze structure of ferroelectric domains (left), corresponding to Figure 2a.

### E. Čerenkov SHG Microscopy

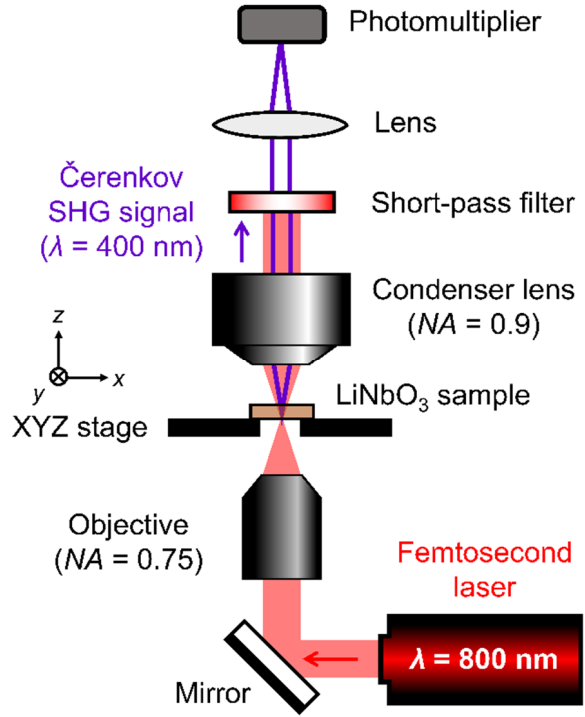

**Figure S8.** Sketch of the Čerenkov SHG microscope employed for 3D ferroelectric domain mapping.

The domain walls in Figure 4 and 6 are detected using a Second Harmonic-Generation (SHG) laser scanning microscope consisting of a mode-locked Ti:sapphire laser (Vitesse, Coherent) and an inverted optical microscope (Nikon eclipse, Ti-U).[65] The setup is schematically depicted in Figure S8. The laser emits laser pulses at 800 nm with 80 MHz repetition rate, 90 fs pulse duration at an average power of 270 mW. The laser beam is expanded and fills the rear aperture of the microscope objective with a numerical aperture of 0.75. The laser power is attenuated by a combination of half-wave plate and polarizer to prevent all-optical domain inversion during the measurement.[41] The maximum power in the sample plane is below 100 mW (1.25 nJ pulse energy). The light is focused into the $z$-cut crystal mounted on a piezo stage (P-545, PI nano) with a $xyz$-travel range of 200 µm × 200 µm × 200 µm and a position resolution of 1 nm. The $1/e^2$ diameter of the diffraction-limited light spot is estimated as $d = 2\lambda/\pi NA = 680$ nm. The crystal is moved through the laser focus in $xyz$-steps of typically 1 µm × 1 µm × 4 µm. The second harmonic light is collected by a condenser lens with a numerical aperture of $NA = 0.9$ and recorded by a photomultiplier (H6780, Hamamatsu) as a function of the focus position. The fundamental wave is effectively blocked by a short-pass filter. The piezo nano-stage is mounted on a *xy*-stage (M-545), which is moved by stepper motors with a travel range of 25 mm × 25 mm and a minimum incremental motion of 1 µm. This allows for scanning of domains with a larger diameter than 200 µm (see Figure 4a).

If the laser focus is located within the LN crystal in a volume without polarization changes, i.e. with a constant $\chi^{(2)}$ nonlinearity, no second harmonic wave is generated. The reason for this lies in the Gouy phase of the focused Gaussian laser pulse, which cancels out any second harmonic, as we are working in a dispersion regime with positive phase mismatch $\Delta k = k_{2\omega} - 2k_\omega > 0$, with the wave vector $k_{2\omega}$ and $k_\omega$ of the second harmonic and fundamental wave, respectively. However, if the laser beam is focused on an interface where the nonlinearity/polarization changes, a second harmonic wave is generated. This happens, for example, at the surface of the crystal, but also when the polarization in the crystal changes from $-P_s$ to $+P_s$ (or from $+P_s$ to $-P_s$), i.e. when the domain wall lies in the *xy*-plane. A second harmonic wave is then generated, which propagates mainly parallel to the *z*-axis. On the other hand, if the laser focus is located in a domain wall that lies in the *z*-plane, a second harmonic wave is generated that propagates at an angle $\theta$ determined by the longitudinal phase matching condition $\cos\theta = 2\,k_\omega/k_{2\omega}$, also known as Čerenkov-type phase matching.[64] To detect second harmonic waves at large Čerenkov angles, a condenser with a high NA is required.

**F. Quasi-Circular Domains under Tighter Focusing Conditions**

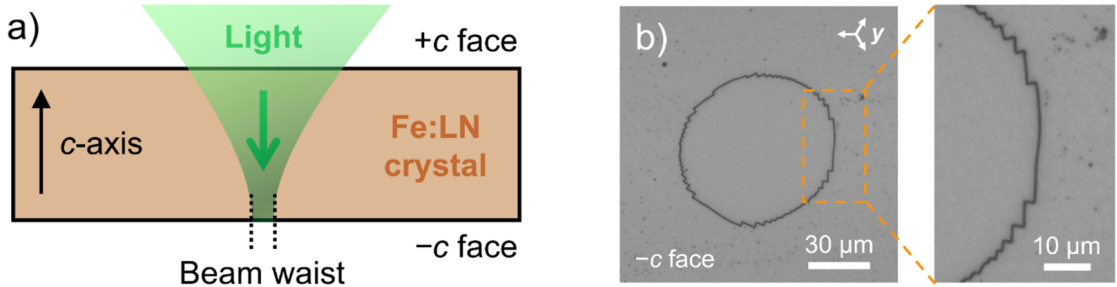

**Figure S9.** a) Schematic diagram of the focusing conditions employed in Section 3.3, with the laser beam waist at the bottom face (the −*c* face). b) Example of a quasi-circular domain obtained with these focusing conditions, using an intensity of 0.42 MW cm$^{-2}$ and an exposure time of 9 seconds. The inset shows how the domain wall is made of multiple sharp steps, with the straight facets preferentially perpendicular to the *y* directions. The location of the longest straight facets coincides with the −*y* sides of the circle.